# A physical model for competition between biological speciation and extinction

Çağlar Tuncay
Department of Physics, Middle East Technical University
06531 Ankara, Turkey
caglart@metu.edu.tr

**Abstract:** Time evolution of number of species (genera, families, and others), population of them, and size distribution of present ones and life times are studied in terms of a new model, where population of each genetic taxon increases by a (random) rate and decreases by (random) division (fragmentation). Each of log-normal, exponential and power law distributions in various empirical data about mentioned events are considered for gradual, intermittent and abrupt evolution cases. Results come out in good agreement with several theoretical literatures and with available empirical data. Sum of exponential functions with independent random amplitude and exponent are shown to be useful to express many aspects of the present competition, and they are shown here to involve power law with power -1.

1     **Introduction:** It is obvious that, there may be no direct observation for biological evolution in past. The empirical data for the past come from excavations, i.e. fossil records; and, these for present come from observation on several living species. Frequently, biological competition between speciation (origination) and extinction at present and in past is expressed in terms of self organized criticality (SOC), since some real data exhibit power law.[1-8] Many other theoretical approaches are also available in literature. (See the relevant text and references in [1-8]) Reviewing the literature (about SOC and discussing other approaches) are kept beyond the scope of this contribution. We present a new model in the following section, where population of each biological taxon increases by a random rate [9] and decreases by division (fragmentation) [10]. We consider also intermittency (where we have random fragmentation) and abrupt extinction (punctuated evolution).

   Our essential aim is to show that, log-normal, exponential and power law distributions in various empirical data may be result of random population growth rate and random fragmentation, which are two opposite effects taking place within the competition between speciation and extinction. Secondly, the historical and present competition between speciation and extinction may be shaped by the relevant parameters about population growth rate and fragmentation. Namely, for a given range of the relevant parameters one of the log-normal, exponential and power law, etc. patterns may be formed; and for another combination of the same parameters another pattern may be formed. Thirdly, all of these formations may be emerging from randomness, rather than from (or besides of) a universal driving mechanism; and, the mentioned randomness may be expressed in terms of sum of exponential functions with independent random amplitude and exponent. Applications and results are exhibited in the next section; where, we also introduce and study some properties of some large number sum of the mentioned exponential functions (with random amplitude and random exponent, which are completely independent). These functions are utilized to describe several aspects of the subject; the rates for speciation and extinction, power laws involved, and life time, etc. Last section is devoted for discussion and conclusion.

2     **Model:** The present subject involves many taxa (taxonomic layers; species, genera, families, …, orders) and a huge number of members. Moreover, the population of a species is known to vary from few or one for a species near to become extinct, to $\sim 10^6$ bacteria per milliliter of seawater [11]. The relevant time range covers many million years, and the physical parameters (climate, chemical composition of air and seas, etc.) of the physical domains display variations (fluctuation) from time to time. Moreover, many mechanisms (from molecular level up to global ones, such as climate, and to extraterrestrial ones, such as falls of meteorites, etc.) are known to be effective on evolution. So, there may be no unique mechanism, which governed all of the mentioned events in past, and is still governing these at present. Instead, several different mechanisms might have taken place for several taxa in several time periods. In other words, not a single mechanism (such as SOC) might have been running biological evolution forever. No microscopic (agent based) or macroscopic (taxon based) satisfactory external mechanism could easily be imagined as a unique drive for such a diverse taxa and huge population to self organization and criticality. Yet, SOC (and any other mechanism) might have been effective in some cases. The original model is free from any hypothesis; and the following remarks of us are worth to be underlined:

1) Several external (a-biotic) and internal (biotic) effects might have been together shaping the evolution in a given period of time. We have a limited chance to discover the reasons in past, and we will not consider them in the present contribution; instead we will focus on the results of them. We assume that the environmental conditions governing the number and the populations of a lower taxon within a high taxon may govern the number and the populations of higher taxa in similar way; and, vice versa. And, the same or very similar biotic reasons for abrupt speciation may be valid for abrupt extinction, i.e., mass extinctions may very well have occurred because of biotic reasons, with few probable exclusions. As any individual dies because of pure biotic reasons, any species, genus or higher taxon may become extinct because of similar (pure biotic) reasons. Any taxon may have been adjusted from speciation, to become extinct ultimately (due to its chemical, and physical and biological nature, etc). As a result, we consider extinction as random, as well as we consider speciation as random. Moreover, once the reasons (where, the external ones are random) are kept out of the formalism, their results may well be considered as random.

2) All taxa are equivalent in the following manner: What might be valid for any taxon at any time that may be valid for another taxon at (the same, or) another time. And, biological inheritance and environmental conditions together determine the crucial parameters (life time, population size and population rate, for example) of all taxa.

Within the original model, number of lower taxa for a taxon under the consideration is ($M(t)$), which varies in time ($t$) according to several conditions (environmental ones; climate, food, room, etc., and biotic ones; genetic, physical, chemical, etc.). Each member (lower taxon) ($i$) in the given taxon might have emerged from a common ancestor at $t = 0$; yet, we may consider more than one ancestor equally well, since it amounts only to shifting the origin of time in our model. At present, different members (genes, species, etc.) may display different size distribution (power law, log-normal, etc.) according to their rates of birth and death, population (abundance), and other conditions. During evolution we assign a random ($0 \leq r_i < 1$) population growth rate ($R_i$) to each member, $0 < R_i = Rr_i \leq R$, where $R$ is constant. Thus, $R$ is the population growth rate for a species, which spreads most quickly on the world, i.e., it is the upper bound for the population growth rate for all the creatures as well as it is the upper bound for the rate for an emerging species, at $t$.

We consider evolution in three scopes:

a) Gradual (continuous) evolution; where we have population (abundance) increment and division (fragmenting, splitting) in each taxon at each time step.

b) Intermittent (discrete) evolution; where we have intermittency within random fragmentation and random population increment.

c) Evolution in terms of abrupt (mass) speciation and extinction (punctuation); where we have random and intermittent speciation and extinction events, which are big in size.

What we mean by fragmentation (division, splitting) of a biological taxon is as follows: Through the evolution, each lower taxon, at each time step (with probability *H*, in case of intermittency) gives birth to a new one with the splitting ratio *S* (mutation factor) which is assumed to be the same for all of the lower taxa. For example, one of the environmental conditions may change (specifically, the weather may cool down, say); as a result, some creatures may adapt and survive; some may not adapt and become extinct. The present (and other) change(s) in the ecosystem may prompt new species to emerge. We count the net change within each lower taxon as splitting (fragmentation). Within the present mutation, if the current population of the agent (*i*) is $P_i(t)$, $SP_i(t)$ many members form another lower taxon and $(1-S)P_i(t)$ many survive. And, due to the present splitting, the number of lower taxon, ($M(t)$) increases by one; if two lower taxa split at *t*, then $M(t)$ increases by two, etc. Please note that, results do not change if $1 - S$ is substituted for *S*, i.e., if the mutated and surviving members are interchanged. So, the greatest value for *S* is 0.5, effectively, and total population ($W(t)$) during any splitting remains constant; yet, $W(t)$ may change in time. Secondly, formation of a new lower taxon via the present splitting (mutation) may also be considered as: A new parameter *S'<1*, which is independent of *S*, may be set; and, it may be assumed that, a new lower taxon emerges with a population of $S'P_i(t)$, and a lower taxon loses its population by $SP_i(t)$, independently, or vice versa. (It is obvious that, $W(t)$ does not remain constant during the current extinction and speciation processes, since $S+S'\neq1$. And, loss of population of a taxon is independent of the (average) life time of its individuals.) Also, one new generated taxon may lose population in further time steps; and whenever any taxon has population less than one, and then we consider the situation as (gradual) extinction. Later, we consider also abrupt mass extensions (punctuation), where a taxon with a population bigger than unity, totally becomes extinct instead of splitting.

Furthermore, our time steps follow historical time (some million years, eons, millennia, centuries, etc.), i.e. they do not increase by one at each emergence of a genus or taxon. Thus our model is defined as follows:

*Initiation*: We assume *M(0)* many (lower) taxa existed initially, and each of these may be assumed to have (equal initial population ($P_i(t = 0)$, or) more or less the same population, at least in order of magnitude. Yet, to study the effect of small sizes, we consider random initial populations, and assign $Pr_i$ many (lower) taxa to the taxon under consideration, where *P* is some constant and $r_i$ is a random real number $0 \leq r_i < 1$, with $i \leq M(0)$. The population growth rate ($R_i$) is also fixed initially, and not varied in time. (Further generation lower taxa randomly get new $R_i$ during splitting, and do not change this parameter later.) We consider also the very special case, *M(0) = 1*, within some of our applications.

*Evolution*: We let the lower taxa grow in time, within a multiplicative noise process,

$$P_i(t) = (1 + R_i)P_i(t - 1) , \qquad (1)$$

and if a random number is smaller than the splitting probability *H* the lower taxon (*i*) splits. It is obvious that, *H =1* is for gradual (continuous) evolution. If for *H << 1* this random number is larger than some *G* near 1, the taxon becomes extinct.

Population of (higher) taxon (*W*) is,

$$W(t) = \sum_{i=1}^{M(t)} P_i(t) . \qquad (2)$$

Please note that, the introduced parameters have units involving time, and our time unit is arbitrary. After some period of evolution in time we (reaching the present) stop the simulation and calculate the (cumulative) probability distribution function (PDF) for the number of lower taxa and size, and for other functions such as life time, etc. Number of interaction tours may be chosen as arbitrary, with different time units; and, the parameters (with units) may be refined accordingly. Yet, in most cases relative values (with respect to other cases; population growth rate in different runs, for example) and ratios of the parameters (the ratio of *S* to *H*, for example) are important.

**3  Applications and Results:** The basic parameters are: *M(0)* (initial number of lower taxa, ancestors), $P_i(t = 0)$ (=$Pr_i$, initial population of each lower taxon, and uniqueness ($r_i = 1$, for all *i*) or randomness of it ($0 \leq r_i < 1$, for all *i*)), $R_i$ (= $Rr_i$, population growth rate), and *H* (intermittency factor for splitting, assumed to be the same for all of the lower taxa). The splitting ratio *S* is also considered as universal. Some of the declared parameters would be eliminated if we knew the real historical data. Please note that, *M(0)* and *P(0)* define the origin of our time scale, and an increase in *M(0)* and in *P(0)* means shifting the time origin forward, and vice versa. On the other hand, *R* and *H* are defined per unit of time. So scaling of only one of them means scaling the time axis by the same factor, but inversely, where results remain invariant.

We don't know the life, population (*W(0)*) and number of genetic lower taxa (*M(0)*) at the beginning of world history, and $R_i$ is also unknown. We run our simulations for various *M(0)*, $P_i(0)$, and $R_i$ for the three regimes of gradual (*H=G=*1), intermittent (*H << G = *1) and punctuated (*H << G <* 1) evolution.

In the following paragraphs, we consider log-normal, exponential, and power law distributions for several biologic functions; where, our aim is to discover and exhibit relations between type of distribution that comes out, and the parameters *S*, *P*, *H*, *M(0)*, etc. In fact, numerical precisions within our parameters and number of time tours are not very closely relevant for the results, due to random processes involved. Our essential aim is (section 1) to show that, log-normal, exponential and power law distributions, etc. in various empirical data may be result of random population growth rate and random fragmentation, which are opposite effects within the competition between speciation and extinction, where randomness rather than (or besides of) a universal driving mechanism is crucial.

**3a  Gradual evolution** (*H=G=*1)**:** In a given niche, and for a biological taxon, we suppose that, initially we have *M(0)* lower taxa (species in a genus, genera in a family, etc.), which have random population, with $P_i(0) \leq P$, and random population rate, $R_i \leq R$. We suppose further that, population of each lower taxon increases at each tour, and each lower taxon splits by a factor of *S*, at the same time.

**3a-i  Log-normal distribution for size:** In Figure 1a, we display the (cumulative) size distribution at the end of 1000 tours, with *M(0)*=100, *P*=1.0x10$^5$, *R*=3.0x10$^{-3}$, and *S*=1.0x10$^{-3}$; so we have population growth rate and splitting factor in the same order of ten. At the end, we have an organization within the niche that, the ancestor taxa survive with high population and we have many new speciation (more in number) with smaller populations (left), where the effect of mutation (fragmentation) is clearly observable. Distribution for the generated lower taxa is slightly asymmetric log-normal. Mammals and birds are found to depict log-normal size distribution. (For slightly asymmetric log-normal size distribution for mammals and birds, and also for the languages at present; see, [12- 20]) It is obvious that, not all of the mutations are successful, and whenever the population of a taxon becomes less than unity, we consider it as an extinction event. In Figure 1b, we display the ratio of extinction events to the number of current taxa, where it may be observed that, each taxon has the chance to give birth to an unsuccessful mutation, as time goes on; and, the inset is for the size distribution of extinction events, where linearity is clear.

Now, let's increase $S$ by a factor of ten and keep $R$ as before (not shown). In this case, we have a different picture; where new species quickly originate, also which split in the next time step(s), if they do not become extinct. Total population saturates (i.e., becomes asymptotically horizontal) and we have the size distribution similar to (but not exactly) an exponential one. And due to rapid speciation and extinction, we now have the extinction events following a power law, where the power is almost plus two. If instead we decrease $S$ by a factor of ten and double the population rate, then we have another type of size distribution function, where the number of taxa decreases linearly with logarithm of size for sizes less than 100; and, for higher sizes, we have almost horizontal behavior about ten for the average of number of taxa. Here, number of taxa increases linearly with time, and total population increases exponentially with time. In this case, we again have a power law for total extinction, where the power is about 1.5. On the other hand, we may increase the splitting factor (to 0.49) and have a smaller value for $P$ (=0.0003) (with $M(0)$=1000, and $P_i(0) \leq 10,000$). In long run, we have extinction of all of the ancestors, and size distribution becomes slightly asymmetric log-normal for size from about unity to about 100,000. In this case, our niche becomes rich in species with small populations.

**3a-ii  Power law:** In the previous case, it is natural that the final results are sensitive to $S$ and $P$, since we have population growth and splitting at each time step for each lower taxon. And, this condition hardly gives power law in size distribution. Yet, some portions of PDF may well be viewed as following power law, for some small size domain(s). For example, the decline (right) in log-normal plot of Fig. 1a may be considered as a decay with power law for the size between 100 and 50,000.

**3a-iii  Exponential distributions:** An interesting feature of our niche is that, we have two main groups of lower taxa, where the group of ancestors is dominant in population and, recently generated group of lower taxa is rich in variety. Now, we increase the population growth rate by a factor of ~3, and divide the fragmentation constant by about hundred, all with respect to these in case 3a-i. So, the ratio of $R$ to $S$ is (=0.01/0.00003) about 300. As a result, size domain increases and number of lower taxa decreases, with respect to previous cases. Relevant distribution will be flattened more and more as time evolves, and in long run we may have a homogeneous distribution of taxa for size (ignoring the region for sparse sizes), where we have some local exponential behavior within the distribution for medium run (i.e., in 1000 time steps). Figure 1c shows such an exponential decay for a considerably big variation in size (from several thousands up to several millions), with a very small exponent. Inset of Fig. 1c shows the total distribution (which may be compared with Fig. 1a, and 1b), where it may be observed that the main body of distribution is similar to that in Fig. 1a. It is worth to note that, the cumulative extinction amount is linear in size for all of the cases considered above. Therefore, we may think that, within a wide variety of the crucial parameters used, such as $S$ and $P$; the main features, such as the number of lower taxa and extinction amount distribution for size remain almost the same. So, the type of the process which runs the evolution (rather than the parameters) is decisive here. In other words, we may have different patterns (log-normal, exponential and power law distributions, etc.) for event frequency, size distribution, etc. for the given process of gradual evolution. Secondly, we may have any of these patterns taking place for a considerable wide range of crucial parameters. Yet, the patterns (log-normal, exponential and power law distributions, etc.) may change from one case to another, if the parameters (or the relevant ratios of them) change considerably ($R/S$, from about unity to about 300, as considered within the present paragraph, for example).

**3b  Intermittency** ($H \ll G = 1$)**:** If time steps of the evolution are long enough, we may have more fragmentation per unit time within taxa. If on the other hand, time steps are considerably shorter, intermittency may become crucial. Moreover, there may be various other reasons for waiting periods of time in fragmentation, as well as in population growth. Now we assume an intermittency factor ($H$) for splitting, which is taken same for all of the lower taxa. Please note that, $H =1$ gives gradual evolution.

**3b-i   Log-Normal distributions:** In Figure 2a, we display our results for $M(0)$=1000, and $S$=0.499; i.e., we have high splitting ratio; where, we also give the time evolution of the size distribution: The thin line represents the given biological taxon at an early time (t=100), thick line represents the same distribution at an intermediate time (t=2500), and the thick squares are for the present time (t = 10,000). $H$ is equal to 0.0004 in all. Please note that, the present distribution is slightly asymmetric log-normal, as the parabolic fit (dashed line) implies. (See, [12- 20]) The extreme ends (for sparse (left) and crowded (right) taxa) approach to unity as $M(0)$ (i.e., the number of ancestors at $t$=0) gets smaller and smaller (we tried $M(0)$ =1000, 100, 10, and 1; not shown.) (See, Fig. 1 in page 277 in [19]). In Figure 2a, we vary $S$ (keeping all of the other parameters same as before), where it may be observed that for very small mutation factor ($S$=0.001) we have a group of small sized lower taxa (open circles), and population of middle (and afterwards smaller) sized taxa increases, as $S$ increases. Thus, distribution becomes log-normal (squares).

Within the present intermittent evolution, if the population of a taxon becomes less than unity, the taxon is counted as extinct. Figure 2c is the relative extinction (i.e., number of extinction per number of lower taxa) in time, where the number of events and extant abundance are same. Secondly, amplitudes decrease exponentially with time, which may be considered as linear (approximately) since the exponent is small. Thirdly we do not have cascades, yet we have many series of extinctions, all of which decay with time. (Or, each series may be considered as a cascade (avalanche), equally well.) Finally, we have a delay period to have the first series of extinctions, which is not random since a long period of time must pass to have the population of a lower taxon to become less than one due to fragmentation.

**3b-ii   Exponential decays:** In Figure 2d we display an exponential size distribution for intermittent fragmentation; where the relevant parameters are: $M(0)$=100, $P_i(0) \leq 10,000$, $S$=0.499, $H$=0.0001, $R_i \leq 7.0 \times 10^{-4}$. Comparison of the present case and the previous one may show that, exponential decay may form in size distribution for relatively small fragmentation and high population rate, and vice versa.

**3b-iii   Intermittent fragmentation and power law:** Figure 2e displays a power law, where the slope of a linear fit is about -1. So, we have $M(P) \propto P^{-1}$, where $P$ is size and $M$ is the present number of lower taxa within niche. A crucial point is that, the power remains almost constant, when the parameters $H$, $P$ and $M$ are changed in considerable amounts. (See, the caption of Fig. 2e, for the values.) It may be mentioned that, the percentage of extinct taxa displays series, the amplitude of which may be considered as decreasing linearly in time for some period; and later it becomes almost constant. Therefore, we have log-normal, exponential and power law distributions, etc. for event frequencies, size distributions, etc. for the intermittent evolution as well as for the case of gradual evolution (as considered in section 3a), yet with different parameters (and their ratios).

**3c   Abrupt (mass) extinction (punctuated evolution)** ($H << G < 1$)**:** We have the parameter $G$ for abrupt mass extinction with the following meaning: If the random number, which is utilized for fragmentation, is greater than $G$, we have total extinction of the relevant taxon. In other words, if the random number defined for each lower taxon for each time step is between zero and $H$ (in section 3a, $H$=1; and in section 3b, $H$<1) then we have fragmentation, with the splitting factor $S$; if the mentioned random number is between $G$ and one, then we have total extinction of the lower taxon; and for an intermediate value (i.e., between $H$ and $G$) the taxon grows in population with rate $\leq R$. It is worth to underline that, increasing *H*, *G*, and *R* means increasing the number of lower taxa, and increasing the amount (abundance) of extinct species (i.e., fossil data), respectively. It is obvious that, there is a critical value for *G* (depending on the number of tours (*t*) and the current number of lower taxa, i.e. indirectly to *H*); where, for $G_{critical}<G \cong 1$ the niche survives, and for smaller values of *G* (i.e., if $G \cong G_{critical}$) the niche may become extinct totally. We predict $G_{critical} \leq 0.9992$, for $H=0.0001$ and for 10,000 tours. Within the following paragraphs we take 1000 ancestors, with random population $\leq 10,000$; and we consider several combinations of parameters, where we define high punctuation as the case for *G* close to $G_{critical}$, low punctuation for *G* close to unity, and intermediate one for $G_{critical}<G<1$. It is obvious that, high punctuation depends on several other parameters as $G_{critical}$ depends on them; and closeness of G to $G_{critical}$ is relative as well as its closeness to unity is relative. Yet, the effect of *G* on several relevant functions is absolute. In short, we consider degree of punctuation with respect to the number of extinction events per time (frequency), and the resulting abundance, etc.

**3c-i   High punctuation:** Under an heavy ($G_{critical} \leq 0.9992$) and a long term (a-biotic, say) effect, due to radioactivity of volcanoes, or poisoning a lake by industry for example, all of the species within the given niche may become extinct in time, whatever the other parameters are. In Figure 3.a, we display mass extinction frequency for $G=0.9992$ (with $H=0.0001$, as before) and inset shows the fossil amount that comes out, where abundance depends on $R(=1.0 \times 10^{-6})$. Since all the members of the given niche is now within fossil data, then we may consider the plot of the inset as the cumulative time distribution of the lower taxa, where we have a power law, with power $\cong 1.7$.

**3c-ii   Intermediate punctuation:** For an intermediate $G=0.9999$, some taxa survive (for $H=0.0001$ and $S=0.499$, and $R=3.0 \times 10^{-6}$). We have homogeneous distribution of punctuation frequency about unity (with $H=0.0001$, as before); and the number of lower taxa fluctuates (about the initial value, which is the number of ancestors, i.e., $M(0)=1000$) as Figure 3.b shows, where the inset is for the size distribution of lower taxa, with a power law 2.

**3c-iii   Intermediate punctuation versus intermediate and high population rate:** Within the previous two cases, we considered low population rate, since we supposed that, populations do not grow, under the present (high) punctuation effect. We now suppose that, populations grow with $R=9.0 \times 10^{-4}$ (i.e., we increase the rates by a factor of 300). Number of lower taxa still fluctuates, yet total population grows exponentially (with $H=0.0001$, as before). Size distribution of taxa is now an exponential decay for a wide range of (large) sizes. Time distribution of mass extinction displays two different behaviors: it is power law for small $t<4000$, and afterwards it becomes an exponential growth.

For higher growth rates (up to $R=1.2 \times 10^{-3}$ or so, where the total population explodes), we do have unimportant variations within the relevant functions; time distribution of mass extinction becomes exponential for smaller *t*, for example.

For *rare (light) punctuation*, where *G* is very close to unity (and extinction abundance may be omitted), other parameters gain importance and we have the cases that we studied previously (i.e., we have the patterns in growths in population and number of taxa, size distribution, etc. similar to the ones, which we have for intermittent and gradual cases).

**3c-iv   Punctuation and log-normal distribution:** We have (random) elimination of taxa under punctuation; so, we may have log-normal size distributions under punctuation, if punctuation is not so strong to disturb the present log-normal distribution. $G=0.9998$ is such an intermediate value to have a log-normal distribution for 1000 ancestors with random population $\leq 10,000$ at a time which is 1000 time steps before the present time, and with $S=0.499$, $H=0.002$, $R=1.0 \times 10^{-4}$. In Figure 3c we display the effect of increasing punctuation on the log-normal size distribution; for $G$=0.9998, 0.999, 0.995 as light, intermediate, and heavy punctuation, respectively. For the present case (with, $H$=0.002) we predict ($G_{critical} \leq 0.995$). For light punctuation ($G$=0.9998), population of the taxon decreases (almost) linearly with time, number of lower taxa increases exponentially (with small exponent); as a result, number of extinct lower taxa per number of current lower taxa decreases, and the extinct cumulative abundance increases (almost) linearly in time. For intermediate punctuation ($G$=0.999), population of the (higher) taxon decreases exponentially (with small exponent), and the number of lower taxa increases exponentially (with a smaller exponent, or almost linearly), but the extinct cumulative abundance slowly increases in time, and it becomes saturated (asymptotically horizontal). For high punctuation ($G$=0.995), population of the (higher) taxon decreases exponentially, and the number of lower taxa turns out to be an exponential decrease; as a result, extinct abundance per time decreases exponentially with time and the cumulative extinct abundance saturates more quickly. Please note that, the number of lower taxa increases for light and intermediate punctuation and it decreases for heavy punctuation. It is because G+H becomes unity in between the intermediate and high punctuation cases, where we have fluctuation in the number about $M(0)$, i.e., the number of lower taxa neither increases nor decreases (but fluctuates) for $G+H$=1, as we had seen similar cases within previous paragraphs (see, Fig. 3b and related text). In Fig. 3c it may be observed that, effect of punctuation is global on niche.

**3d   Finite, yet big number sum of exponential functions with random and independent amplitude and exponent:** Many aspects of the present competition between extinction and speciation of species may be considered in terms of exponential functions ($y_{ij}(x)$), where the amplitude and exponent are random and independent; for a real number $b$,

$$y_{ij}(x) = A_i \exp(B_j bx)  .  \qquad (3)$$

In Eq. (3) $A_i$ is one random number $A_{min} \leq A_i \leq A_{max}$ and $B_i$ is another random number $B_{min} \leq B_j \leq B_{max}$ which are independent of each other. In other words, we select $A$ and $B$, from different sets of random numbers, between which there is no connection. We define,

$$Y(x) = \sum_i^I \sum_j^J y_{ij}(x)/IJ  ,  \qquad (4)$$

with some big yet finite $I$ and $J$, where the order of sum is not important due to independence of the random numbers.

It is known that many physical functions show exponential decreasing distributions and exponential decays in time, and we will take $b$ in Eq. (3), (which is indeed a scaling factor for $x$-axis), as negative for positive $B_j$, ($b \rightarrow -b$), thus we will have negative exponents. Figure 4a is for $y_{ij}(x)$ (Eq. (3)) with b=0.1 and $A_{min}= B_{min}$ =0.0,  $A_{max} = B_{max}$=1.0, which may be considered as the size distributions for taxa within niche, for taxa within fossil data, (population of cities, number of speakers of human languages), etc. The inset is for the evolution of the cumulative sum of $y_{ij}$, i.e. the integral $f(x)=\int^x y_{ij}(x')dx'$, which is time evolution of fossil cumulated within niche (after $x' \rightarrow t'$, and $x \rightarrow t$) if $y_{ij}$ is utilized for the time evolution of total population (of the taxa) within niche. (If one ignores the portion for $t$<2,000, with arbitrary unit for time, or considers time steps longer than 2,000; the present information and the plot of the inset will appear as linear in time.) Equally well, $y_{ij}(x)$ may be considered as population of each lower taxon ($x$). Then, $f(x)$ becomes the probability distribution (PDF) of cumulative population over the size. In this case, a rank plot (reorganization of data for $y_{ij}$, say in ascending manner) shows that, about 7,000 lower taxon has population less than $10^{-40}$, (amplitudes are between 0 and 1, and we can not determine a limit for a lower taxon to become extinct, which is unity in reality) so we have very heavy punctuation (not shown). It is worth to note that, the mentioned rank plot (or, rank/frequency plot; see, [21,22]) gives a power law for a very wide range of size, i.e., from $10^{-40}$ to about unity and PDF for size distribution (not shown) is similar to the one in Fig. 3c (the lowest plot). On the other hand, plot of the inset in Fig. 4a may be considered as linear in $x$ for $2{,}000 < x$, where the slope is very small (the plot may also be considered as an exponential growth with very small exponent, naturally).

For the sum in Eq. (4), one may select the random numbers same for all $x$, and one may select new (different) random numbers at each $x$, where all of the random numbers are completely independent ($A_i$ is one random number, $A_{min} \leq A_i \leq A_{max}$; and $B_i$ is another random number, $B_{min} \leq B_j \leq B_{max}$) in all. Figure 4b displays $Y(x)$ (Eq. (4)) with $I=J=1000$, for $A_{min} = B_{min} = 0.0$, $A_{max} \leq 1.0$, $B_{max} \leq 1.0$, where we have no fluctuation in $Y$ for the random numbers selected same for each $x$ (dashed thick line) and fluctuation in $Y$ increases in magnitude, as $x$ increases (fluctuating plot) for the random numbers (re-)selected for each $x$. And, in both cases we have power law; as the arrow with slope -1 indicates in log-log axes.

Obviously, the important issue in Fig. 4b is that $Y(x)$ is a power function for large $x$ with the power of (about) minus unity, i.e.,

$$Y(x) = \sum_i^I \sum_j^J y_{ij}(x)/IJ \propto x^{-1} \quad , \tag{5}$$

which is similar to Pareto-Zipf law, if $x$ stands for city population. We may state that, if the probability of finding a city with population about $x$ (say, $x \pm dx$) decreases exponentially with independent random amplitude and exponent, then the sum of mentioned probabilities gives (nearly) the famous power law. In other words, reason for the mentioned power law is the randomness in amplitude and (negative) exponent of the probability of finding a city with population about $x$ (say, $x \pm dx$). Similarly, as pointed out in [23] that "existence times of a family within a specified order was exponentially distributed; a fact that indicated the randomness of extinctions as well as the distinctive different life times between the different order". This is known as Van Valen's [24] original Red Queen hypothesis and then life time of species should be distributed following a power law with power (about) minus unity as Eq. (5) implies (after $x \rightarrow t$).

Clearly, Eq. (5) is valid for every situation (competition between cities, that between languages, and similarly between speciation and extinction, etc.), if it satisfies the given conditions (i.e., random and positive amplitudes, independently random and negative exponents). For homogeneously (yet, independent) distributions for the random numbers $A_i$ (for the amplitudes) and $B_i$ (for the exponents), with $A_{min} \leq A_i < A_{max}$ and $B_{min} \leq B_j \leq B_{max}$, one may utilize the general theorem of central limit with large $N$ or just convert the double sum into a double integral over $A$ ($A_i \rightarrow A$) and $B$ ($B_i \rightarrow B$), which vary linearly (since the random numbers are homogeneously distributed) between the extrema to obtain the following equalities:

$$\begin{aligned} Y(x) &= \sum_i^I \sum_j^J y_{ij}(x)/IJ \\ &\propto -(A^2_{max} - A^2_{min})[\exp(-B_{max}bx) - \exp(-B_{min}bx)]/(2bx) \quad , \end{aligned} \tag{6}$$

and, for $B_{min} = 0$, we have

$$Y(x) \propto -(A^2_{max} - A^2_{min})[\exp(-B_{max}bx) - 1]/(2bx) \quad , \tag{7}$$

which may be further simplified, for $1 \ll 2bx$;

$$Y(x) \propto [(A^2_{max} - A^2_{min})/2b] \, x^{-1} \quad , \tag{8}$$

since, the exponential term approaches to zero as *x* increases. And the result is power law, with power minus one (Eq. 5). It is obvious that, $B_{max}$ may always be taken as unity, and *b* may be varied accordingly. We may state that, our analysis (Eqs. (5)-(8)) is general for any set of independent random numbers for amplitudes (provided $A_{max}$ and $A_{min}$ is finite) and that for exponents (provided $B_{min}$= 0, where $B_{max}$ =1 may be taken after changing *b* accordingly). The inset in Fig. 4b shows $Y(x)$ in Eq. (4) as a thick line and this in Eq. (8) as a dashed line, all with $A_{min}$ =$B_{min}$= 0, $A_{max}$ = $B_{max}$ =1, and *b*=0.1.

**3e    Life times:** For life times, we may simply subtract the number of time step at which a lower taxon emerged, from that one at which the given lower taxon became extinct. In the present section we consider various cases, all of which involve 1000 ancestors with random population (≤10,000); where we consider 10,000 time steps with $R=3.0 \times 10^{-4}$, $S=0.499$, $H=0.0004$, and we have punctuation with various intensity. (We had considered several effects of punctuation in section **3c-i**.)

Figure 5a is life times for lower taxa which become extinct in terms of fragmentation and high punctuation $G=0.99945$ (please note that, $H+G <1$). The inset is probability distribution of life time, where vertical axis indicates number of lower taxa that survived between numbers of time steps given on the horizontal axis. For example, about 10,000 many taxa survived (between 0 and) 1 time step, and less than 10 taxa existed in (between 4 and) 5, and in (between 5 and) 6 time steps. The arrow within the inset indicates that, the distribution is exponential with exponent (about) -0.5. It is a crucial issue that, life times for the current taxa (these, which are present at the end of 10,000 time steps) displays a different picture, as given within Figure 5b, where we observe that distribution is exponential, and, many of the living lower taxa (about 30 %) survive for less than 1000 time steps. Only few (3, as indicated by the point for t=10,000) of 1000 ancestors reach the present time, and rest become extinct on the way or split. So, the maximum life time is the age of ancestors (t=10,000). We have similar results for intermediate ($G=0.9996$, with $1=H+G$.) and light ($G=0.99998$, with $1<H+G$) punctuation cases; in all of which we have exponential distribution (decreasing) of life time for extinct lower taxa with few (less than 10) time steps at maximum. For living lower taxa, life time distribution is exponential in all, and number of ancestors, which are living, increases as *G* increases, where (as considered within previous paragraphs) number of the living lower taxa also increases (exponential in time) as *G* increases. We have about 50 lower taxa which survive about 10,000 time steps for $G=0.9996$. It is worth to note that, distribution of age over the living taxa condenses along a straight line (i.e., dispersion in Fig. 5b becomes more acute) as *G* increases. And cumulative PDF for ages of living taxa (Figure 5c is for $G+H=1$) gives power law at tail, where the power exponent decreases as *G* increases.

**4    Discussion and Conclusion:** As the recent predictions (within the previous paragraph) indicate, we may state that:

    1) Life times for lower taxa which became extinct is exponential, with few time steps (less then 10) for maximum duration.

    2) Life times for the living lower taxa (ages) is exponential, with about 10,000 time steps for maximum age.

    3) The present exponential behaviors originate from randomness: we have arbitrary number of ancestors with random (initial) populations, we have random population rates and random fragmentation (with intermittency, and a universal splitting factor), and we have random punctuation of lower taxa; and finally we obtain exponential longevities for extinct and living taxa.

4) Any lower taxon either becomes extinct within few time steps (less than 10) after its origination or survives much longer. Therefore, the second remark of us, (which is underlined) in section 2, has the following meaning: Not all of the lower taxa of a given (higher) taxon are equal in absolute meaning; since, some of the originated ones become extinct quickly and some others live much longer. Yet, this is valid for all taxa; namely, some lower taxa of each (higher) taxon become extinct quickly, and some others live long.

5) Maximum age of the living taxa is determined basically by the (maximum, or average of the) ages of ancestors, many (or better, some) of which lived till present.

It is worth to note that, within our model, population of each lower taxa increases with a random rate, and some of them split randomly (involving intermittency) and some others become extinct totally with a random rate of punctuation. These two opposite effects drive the system. The situation is similar to a granule in a sand pile, which is in a meta-stable state under the effect of gravitational force and frictional force, which are opposite to each other. It is known that, kinetic co-efficient of friction is smaller than static co-efficient; and, if due to some reason, any sand granule is hit, the co-efficient of friction changes from static one to kinetic, because of the initial speed, which may be small yet it is non zero. And the granule gains kinetic energy. It may hit others, and make them move. As a result, we have avalanche. Up to us, there is no self organization within sand avalanches. (Not all the granules in bulk and even all of these on the surface are organized.) But, we obviously have order (regularity) in terms of exponential life time and power law (which are emerging out of the fundamental randomness) within size and age distribution of living taxa; i.e., within the competition between biological speciation and extinction.

Another crucial issue is that, we may have several patterns (log-normal, exponential and power law distributions, etc.) for each of the several quantities (population, size distribution, longevity, etc.) about various taxa, which may evolve gradually, intermittently, and abruptly, where several combinations of relevant parameters (population growth rate, and factors for fragmentation, intermittency and punctuation; and their ratios) are decisive.

**Acknowledgement**
The author is thankful to Dietrich Stauffer for his discussions about Eq. (6) and contribution here.

**REFERENCE LIST:**

[1]     M. E. J. Newman, Self-organized criticality, evolution and the fossil extinction record. Proceedings of the Royal Society of London B263, 1605-1610 (1996).
[2]     M. E. J. Newman, A model of mass extinction. Journal of Theoretical Biology 189, 235-252 (1997).
[3]     M. E. J. Newman, Evidence for self-organized criticality in evolution. Physica D107, 293-296 (1997).
[4]     M. E. J. Newman, and R. G. Palmer, Modeling Extinction, (Oxford University Press, 2002).
[5]     M. E. J. Newman, SIAM Rev. 45, 167 (2003).
[6]     K. Sneppen and M.E.J. Newman, Coherent noise, scale invariance and intermittency in large systems, Physica D 110, 209-222 (1997).
[7]     B. Drossel, Biological evolution and statistical physics, Advances in Physics, Vol. 50, No. 2, 209-295 (2001).
[8]     D. Chowdhury, D. Stauffer, Evolutionary ecology in-silico: Does mathematical modelling help in understanding the "generic" trends?, J. Biosci. (India) 30, 277 (2005). arXiv:q-bio.PE/0504020
[9]     H.A. Simon, Models of Man, Wiley, New York 1957
[10]    V. Novotny and P. Drozdz, Proc. Roy. Soc. London B267, 947, (2000).
[11]    M. Breitbart and F. Rohwer, Trends Microbiol. 13, 278 (2005).
[12]    J. A. Gauthier, Saurischian monophyly and the origin of birds. Mem. Calif. Acad. Sci. 8, 1–47 (1986).
[13]    L. M. Chiappe, The first 85 million years of avian evolution. Nature 378, 349–355 (1995).
[14]    D. Stauffer, C. Schulze, Microscopic and macroscopic simulation of competition between languages, Physics of Life Reviews 2, 89–116 (2005).
[15]    North American breeding bird survey URL: http://www.mbr.nbs.gov.
[16]    R. H. MacArthur. On the relative abundance of bird species. Proc. Nat. Acad. Sci. USA, Vol. 43, pp. 293–295 (1957).
[17]    T. Gisiger, Scale invariance in biology: coincidence or footprint of a universal mechanism?, Biol. Rev. 76, pp. 161 (2001).
[18]    Ç. Tuncay, A new model for competition between many languages, IJMPC (2007), arXiv:physics/0612137.
[19]    W. J. Sutherland, Parallel extinction risk and global distribution of languages and species, Nature, vol 423, 276-279 (2003).
[20]    I. Volkov, J. R. Banavar, S. P. Hubbell, and A. Maritan, Neutral theory and relative species abundance in ecology, Nature, vol, 424, p 1035-1037 (2003).
[21]    R. Günter, B. Chapiro, and P. Wagner, "Physical complexity and Zipf's law", International Journal of Theoretical Physics 31, 524 (1992).
[22]    R. Günter, L. Levitin, B. Schapiro, and P. Wagner, "Zipf's law and the effect of ranking on probability distributions", International Journal of Theoretical Physics 35, 395 (1996).
[23]    S. Bornholdt, K. Sneppen, and H. Westphal, Longevity of orders is related to the longevity of their constituent genera rather than genus richness, arXiv:q-bio.PE/0608033.
[24]    L. M. Van Valen, Evol. Theory 1, 1-30 (1973).

**FIGURES:**

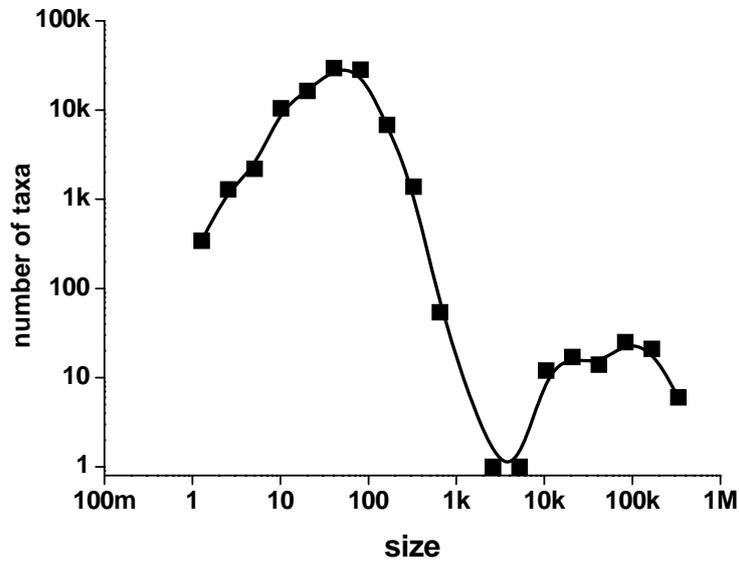

**Figure 1a**    Size (cumulative) distribution of a genetic taxon; with $M(0)=100$, $P=1.0 \times 10^5$, $R=3.0 \times 10^{-3}$, and $S=1.0 \times 10^{-3}$. Please note that, the final distribution (left) is slightly asymmetric log-normal.

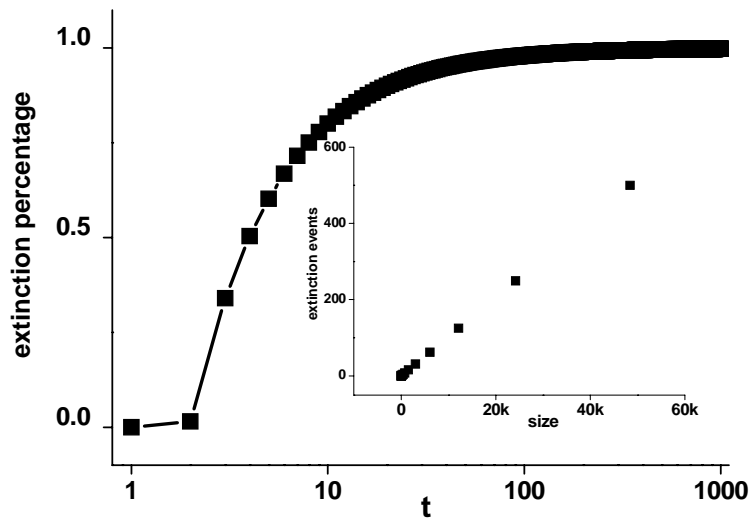

**Figure 1b**    Ratio of extinction events to the number of current taxa, where it may be observed that, each taxon has a chance to give birth to an unsuccessful mutation, as time goes on; and, the inset is for the size distribution of extant, where linearity is clear.

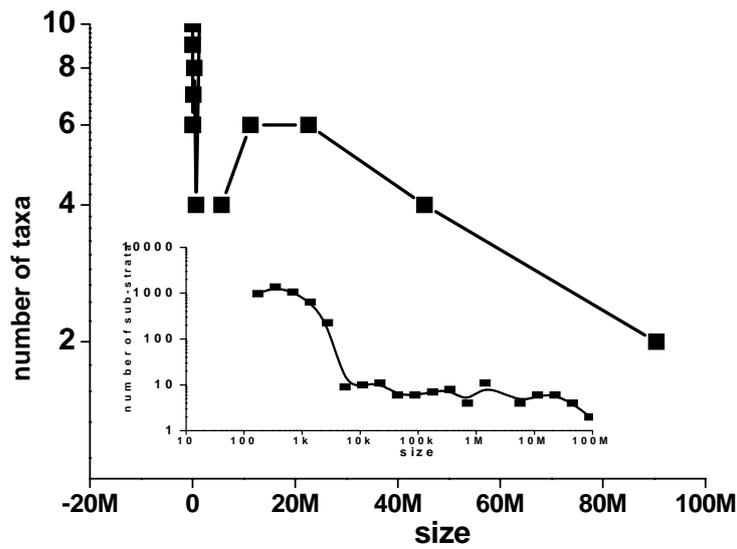

**Figure 1c**   Size (cumulative) distribution of a biological taxon, where the ratio of *P* to *S* is (=0.01/0.00003) and other parameters are as in Fig. 1a. Please note that, some portion of the distribution (inset) may be considered as exponential decay.

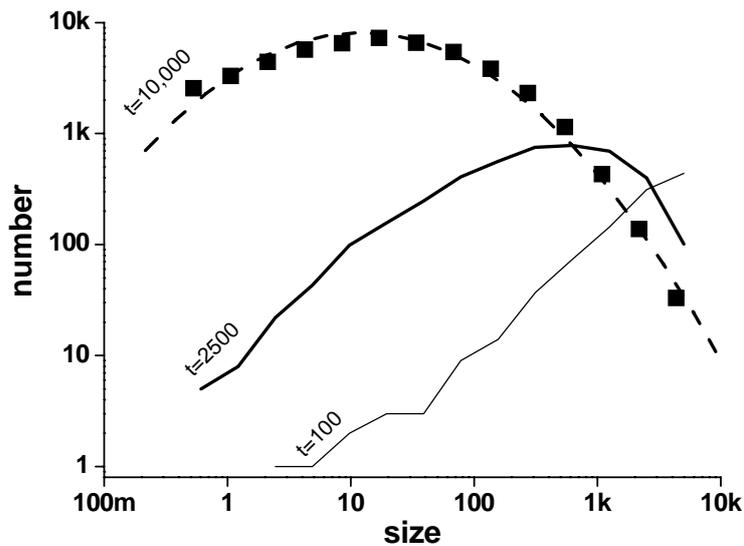

**Figure 2a**   Size distribution (and evolution, where the corresponding period of time is given on the left end of each plot) of a taxon within a given niche; with $M(0)$=1000, and $S$=0.499. ($P$=10,000, and $R$= $10^{-8}$.) Please note that, the final distribution (for the present time, t = 10,000) is slightly asymmetric log-normal. *H* is equal to 0.0004 in all.

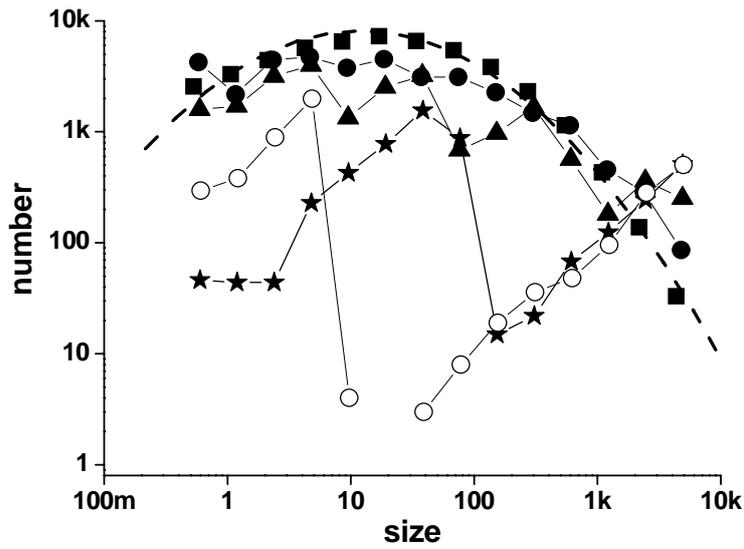

**Figure 2b**   Size distribution for $S=0.499$ (square), 0.2 (full circle), 0.1 (triangle), 0.01 (star), and 0.001 (open circle). All of the other parameters are same as in Fig. 2a. Please note that, for very small mutation factor ($S=0.001$) we have a group of small sized lower taxa, and as $S$ increases population of middle sized lower taxa increases; thus, distribution becomes lognormal (from open circles to squares, which is same as the plot of Fig. 2a).

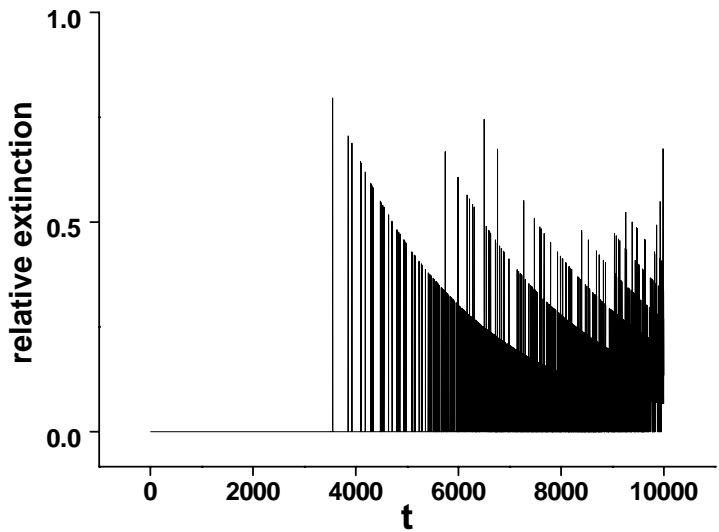

**Figure 2c**   Relative extinction (per population of lower taxa) as a function of time, where the parameters are same as in Fig. 2a. (Unit for the vertical axis is 0.003.)

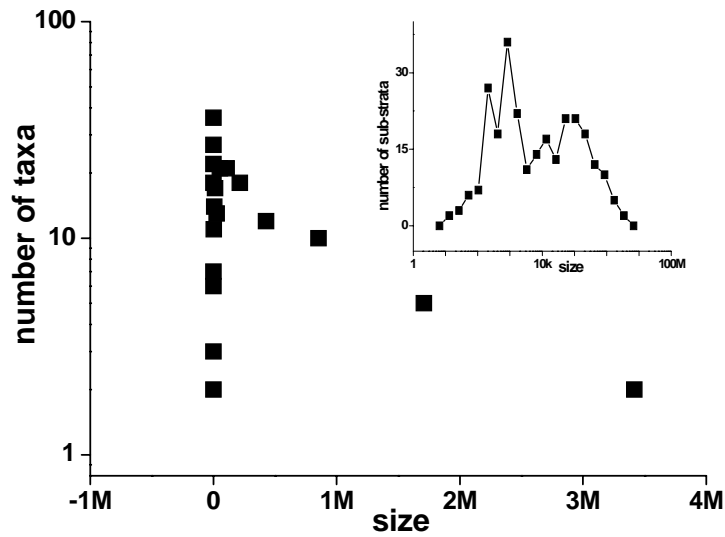

**Figure 2d**  Exponential size distribution for intermittent fragmentation; where the relevant parameters are: $M(0)=100$, $P_i(0) \leq 10,000$, $S=0.499$, $H=0.0001$, $R_i \leq 7.0 \times 10^{-4}$. Inset is the same distribution in linear vertical axis and logarithmic horizontal axis.

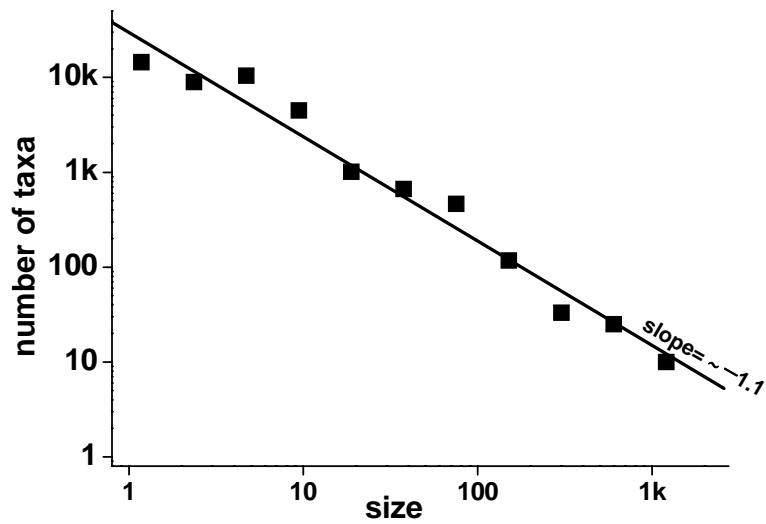

**Figure 2e**  Power law in size distribution for an intermittent evolution. Slope of the straight line (which is a linear fit) is ~-1.1, and it may be considered as -1. Relevant parameters are: $S=0.1$, $H=0.002$, $R=0.00001$, and $M(0)=100$, with $P_i(0) \leq 10,000$.

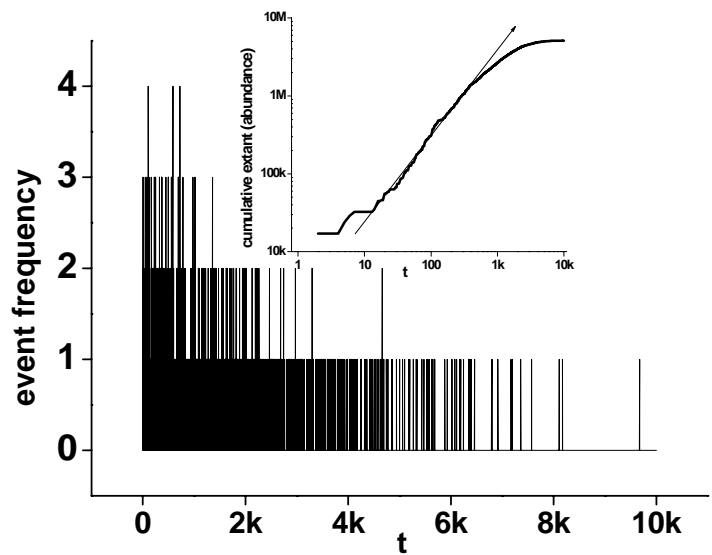

**Figure 3a**   Frequency of mass extinction events (punctuation). Inset is for the cumulative extinction abundance, where a power law is clear, and the power exponent is about 1.7.

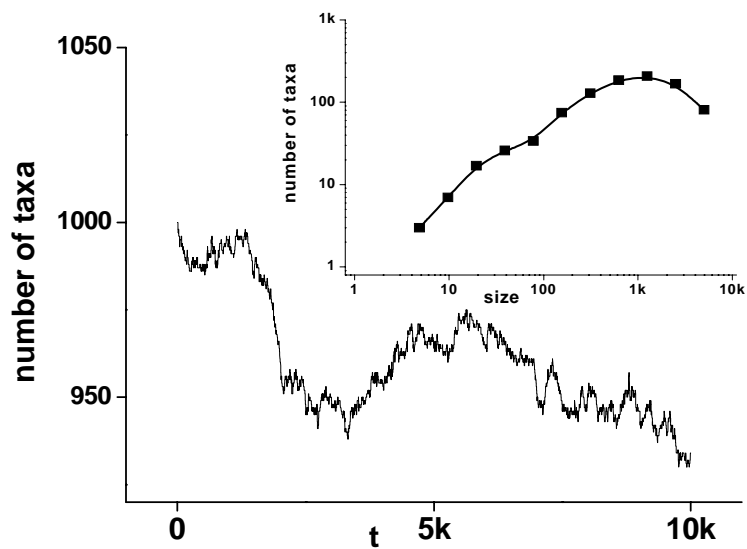

**Figure 3b**   Fluctuation of the number of lower taxa for an intermediate punctuation rate ($G$=0.9999, for other parameters please see the relevant text). Inset is for the size distribution, where we have power law, and the power is positive and about 2.

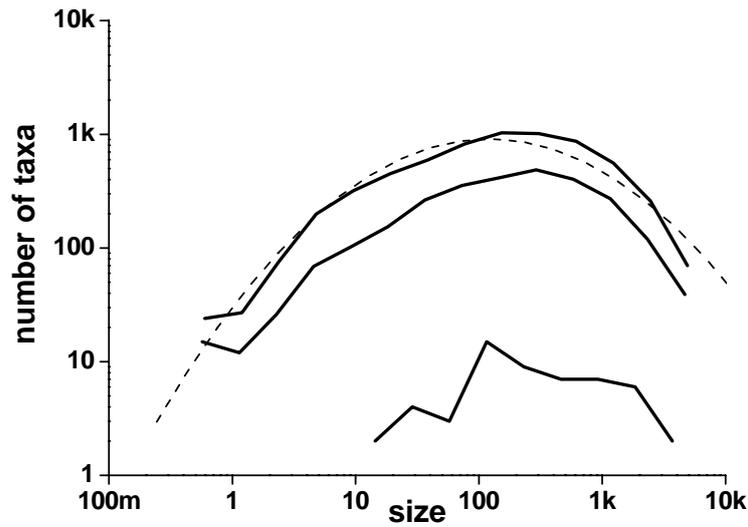

**Figure 3c**    Size distribution of lower taxa under light, intermediate, and high punctuation, from top to bottom. (For the parameters, see the relevant text.) Dashed line is for a parabolic fit, which implies that distribution is slightly asymmetric log-normal.

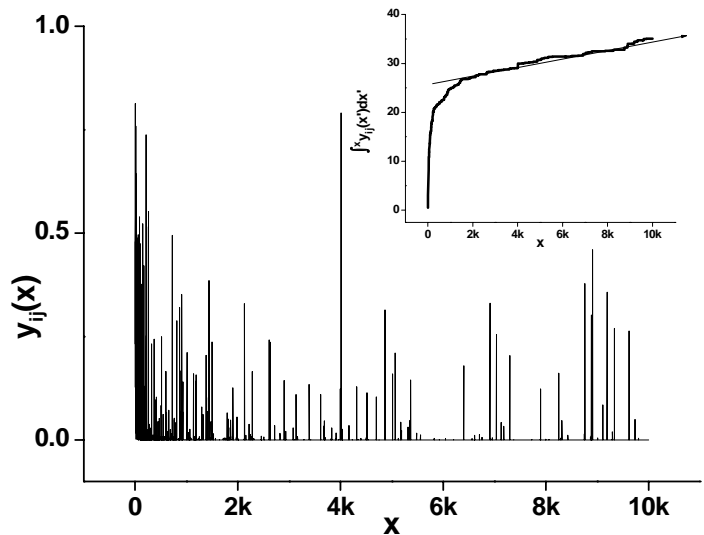

**Figure 4a**    $y_{ij}(x)$ (Eq. (3)) with b=0.1 and $A_{min}= B_{min} =0.0$, $A_{max} = B_{max}=1.0$, which may be considered as the size distributions for taxa within niche, for taxa within fossil data, (population of cities, number of speakers of human languages), etc. The inset is for the evolution of the cumulative sum of $y_{ij}$, i.e. the integral $f(x)=\int^{x} y_{ij}(x')dx'$.

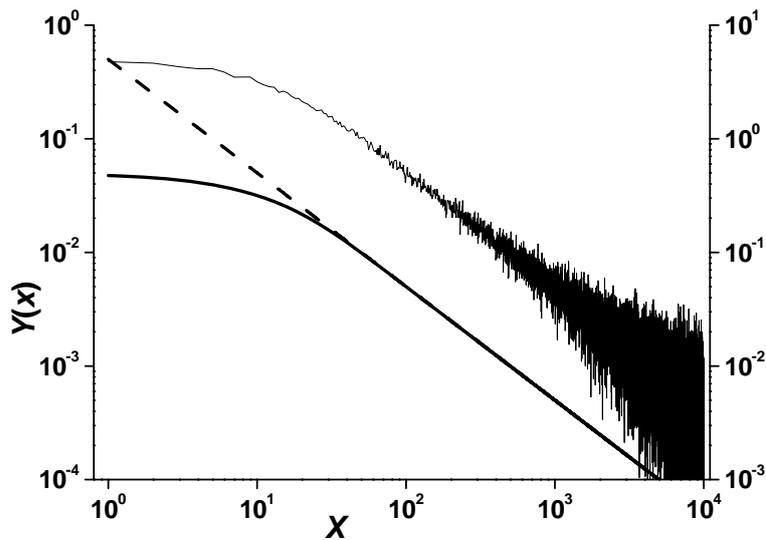

**Figure 4b** $Y(x)$ (Eqs. (4)-(8)) with $I=J=1000$, for $A_{min}= B_{min} =0.0$, $A_{max} \leq 1.0$, $B_{max} \leq 1.0$, where the oscillating plot is for Eq. (4) with the vertical axis on left, and with the vertical axis on right are: The solid plot, for the analytical expression in Eq. (7); and the dashed line, for Eq. (8). Please note that, vertical axes are shifted with respect to each other, yet the units are same. Secondly, the solid and the dashed line has the slope minus unity for large $x$. It is obvious that, the oscillating plot displays power law minus unity (Pareto-Zipf law) for large $x$.

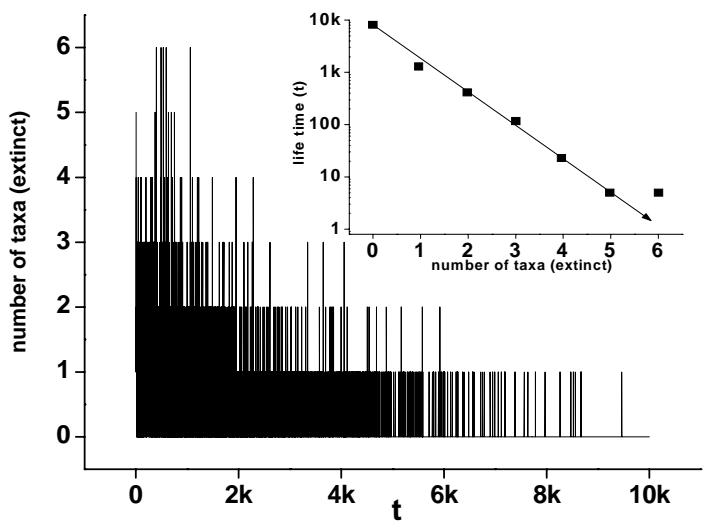

**Figure 5a** Life time for lower taxa which become extinct under high punctuation, where the vertical axis is for the number of lower taxa and the horizontal axis is for their existence period of time. (For the relevant parameters, see section 3e.) The inset displays distribution of longevity over extinct lower taxa, where the plot is exponential with exponent about -0.5.

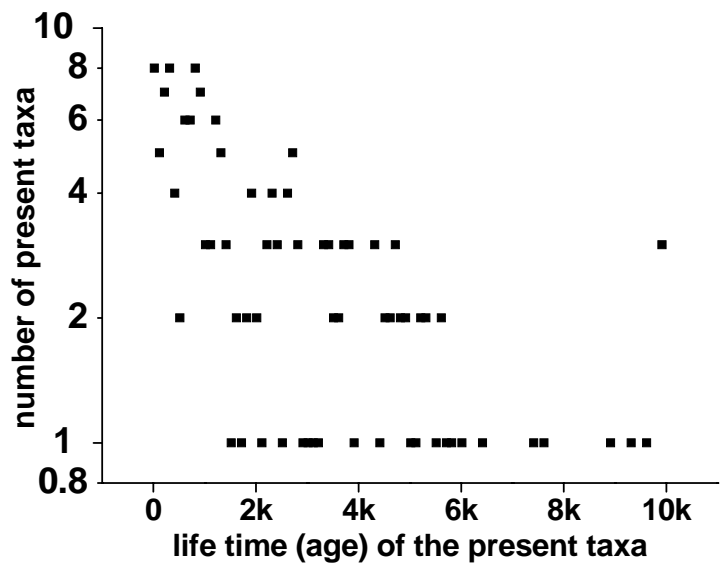

**Figure 5b**  Life time for lower taxa living at present under high-punctuation, i.e., age. (For the relevant parameters, see section 3e.)

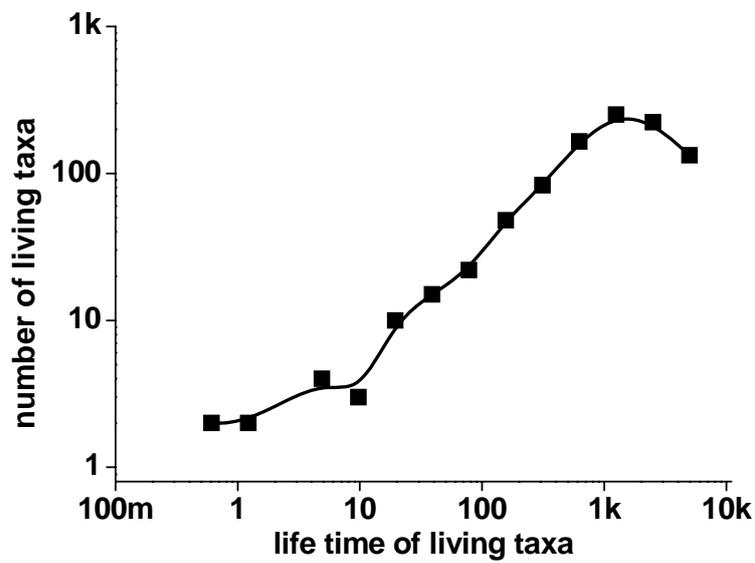

**Figure 5c**  Life time for lower taxa living at present under intermediate-punctuation; $G+H=1$. (For the relevant parameters, see section 3e.)